\documentclass[10pt,a4paper,twoside]{article}
\usepackage{epsfig}
\usepackage{baltlat6}
\usepackage{wrapfig}
\usepackage{dcolumn}
\begin{document}
\ \
\vspace{-0.5mm}

\setcounter{page}{245}
\vspace{-2mm}

\titlehead{Baltic Astronomy, vol.\ts 14, 245--251, 2005.}

\titleb{RADIATIVE TRANSFER PROBLEM IN DUSTY GALAXIES:\\
ITERATION SCALING APPROXIMATION}

\begin{authorl}
\authorb{D. Semionov}{} and
\authorb{V. Vansevi{\v c}ius}{}
\end{authorl}

\moveright-3.2mm
\vbox{
\begin{addressl}
\addressb{}{Institute of Physics,
Savanori{\c u} 231, Vilnius LT-02300, Lithuania}
\end{addressl}
}
\vskip-1mm

\submitb{Received 2005 April 5; revised 2005 May 24}

\begin{summary} We investigate the applicability and accuracy of the
iteration scaling approximation, proposed by Kylafis and Bahcall (1987).
It is shown, that while this method provides results sufficiently close
to the exact solution of radiative transfer problem, care must be taken
in cases when the distribution of interstellar dust significantly
differs from that of light sources.  The problem is successfully
circumvented by using the ratio of scattered light distributions
obtained in the first two iterations.  \end{summary}


\resthead{Iteration scaling approximation}{D.  Semionov, V.
Vansevi{\v c}ius}

\sectionb{1}{INTRODUCTION}

\vskip-2mm

The correct solution of the radiative transfer (RT) problem in the
interstellar medium is very important to a number of astrophysical
topics (e.g., Li \& Greenberg 2003).  However, all algorithms, currently
in use, while being similar in numerical accuracy and applicability,
suffer from a major drawback -- they are quite slow (e.g., Baes \&
Dejonghe 2001; Pascucci \etal\ 2004).  Thus, while the algorithms and
hardware are being perfected, it is important to explore every
approximation that could produce the results with acceptable precision
during computing time which is shorter than that necessary to obtain the
exact solution.  If that approximation introduces errors which are
either below the observational error limit, or are lower than the
intrinsic numerical errors of a given method, then such a solution is
preferable.  In this paper we will test the validity of one such an
approximation for iterative RT problem solving algorithms applied to
spectrophotometric models of dusty disk galaxies.

\vskip-1mm

\sectionb{2}{MODELS}

\vskip-2mm

Kylafis \& Bahcall (1987, hereafter KB87) have proposed an approximate
solution for iterative algorithm, where each iteration accounts for a
single scattering event for all photons in the system. Their method
(hereafter -- iteration scaling approximation, ISA), based on the fact
that for the scattering phase function parameter $g_\lambda > 0.3$
(Henyey \& Greenstein 1941) the scattering is directed mostly forward,
assumes that the ratio of global radiative energy distributions (as
the function of a position within the model and of the direction
relative to the model axes) $I_{j}/I_{j-1}$ for any two subsequent
iterations $j-1$ and $j$ is equal to the ratio of global radiative
energy distribution after the first scattering $I_{1}$ to the initial
distribution of star light $I_{0}$:

\begin{equation}
{I_{j} \over I_{j-1}} = {I_{1} \over I_{0}}.
\end{equation}

\noindent If this approximation produces results of acceptable
accuracy, then the iterative method will have a decisive advantage
over the Monte-Carlo method, since it will be possible to approximately
solve the RT problem up to 10 times faster than is necessary to obtain
the exact solution.

\begin{wrapfigure}{l}[0pt]{6.0cm}
\vbox{\footnotesize
\begin{tabular}{lrrrrrr}
\multicolumn{7}{c}{\parbox{5.5cm}{\baselineskip=8pt
~~~~{\smallbf Table 1.}{\small\ Galaxy model parameters.}}}\\
\tablerule
& S1 & S2 & S3 & S4 & S5 & S6 \\
\tablerule
$z^d_{\rm eff}$, $z_{\rm eff}$ & 0.5 & 1 & 2 & 0.5 & 1  & 2\\
$\tau_V$                       & 1   & 1 & 1 & 10  & 10 & 10\\
\tablerule
\end{tabular}
}
\end{wrapfigure}

To verify the accuracy and applicability of ISA we use the same
models and methods presented in our previous paper
(Semionov \& Vansevi{\v c}ius 2005b). The RT problem is solved
using the Galactic Fog Engine (Semionov \& Vansevi{\v c}ius 2002;
Semionov 2003; Semionov \& Vansevi{\v c}ius 2005a), a code for
spectrophotometric modeling of dusty disk galaxies, employing a
2D iterative ray-tracing algorithm for six disk galaxy model groups
S1--S6, having two values of the optical depth to the model center in
$V$ passband
measured perpendicularly to the disk plane: $\tau_V = 1$
for model groups S1--S3 and 10 for model groups S4--S6. The stars
and the dust follow a double exponential luminosity and mass
distribution,

\begin{equation}
\rho(r,z) = \rho_0 \exp \left( -{r\over r_{\rm eff}} -{\left| z \right|\over z_{\rm eff}}\right),
\end{equation}

\noindent with the same effective scale-length $r_{\rm eff}$ and
scale-heights $z_{\rm eff}$ and $z^d_{\rm eff}$ for the stellar and
dust disks, respectively. The models represent three cases thought
to approximate real disk galaxies -- ``dust within stellar disk''
($z^d_{\rm eff} = 0.5 z_{\rm eff}$, models S1 and S4), ``well-mixed
dust and stars'' ($z^d_{\rm eff} = z_{\rm eff}$, models S2 and S5) and
``dust enveloping stellar disk'' ($z^d_{\rm eff} = 2 z_{\rm eff}$,
models S3 and S6). Optical interstellar dust parameters were
computed using the Laor \& Draine (1993) model approximating the
Milky Way galaxy extinction law.

Each model group consists of M1, M2, M3, M5 and M8 models. For each
M$i$ model eight anisotropic scattering iterations were performed,
the first $i$ iterations being computed exactly, and the remaining
iterations -- using the $i$th order ISA:

\begin{equation}
{I_{j} \over I_{j-1}} = {I_{n} \over I_{n-1}}.
\end{equation}

\noindent The original ISA, as proposed by KB87 (Eq. 1), corresponds
to model M1, while model M8 is computed exactly, without employing ISA.
All models show the same total absorbed energy $E^{\rm abs}_\lambda$
values, differing less than the model numerical accuracy. The energy
defect after the 8th iteration for all models does not exceed 1\% of
the total radiative energy $E^{\rm tot}_\lambda$.

\vskip-1mm

\sectionb{3}{RESULTS AND DISCUSSION}

\vskip-3mm


\begin{figure}
\centerline{\psfig{figure=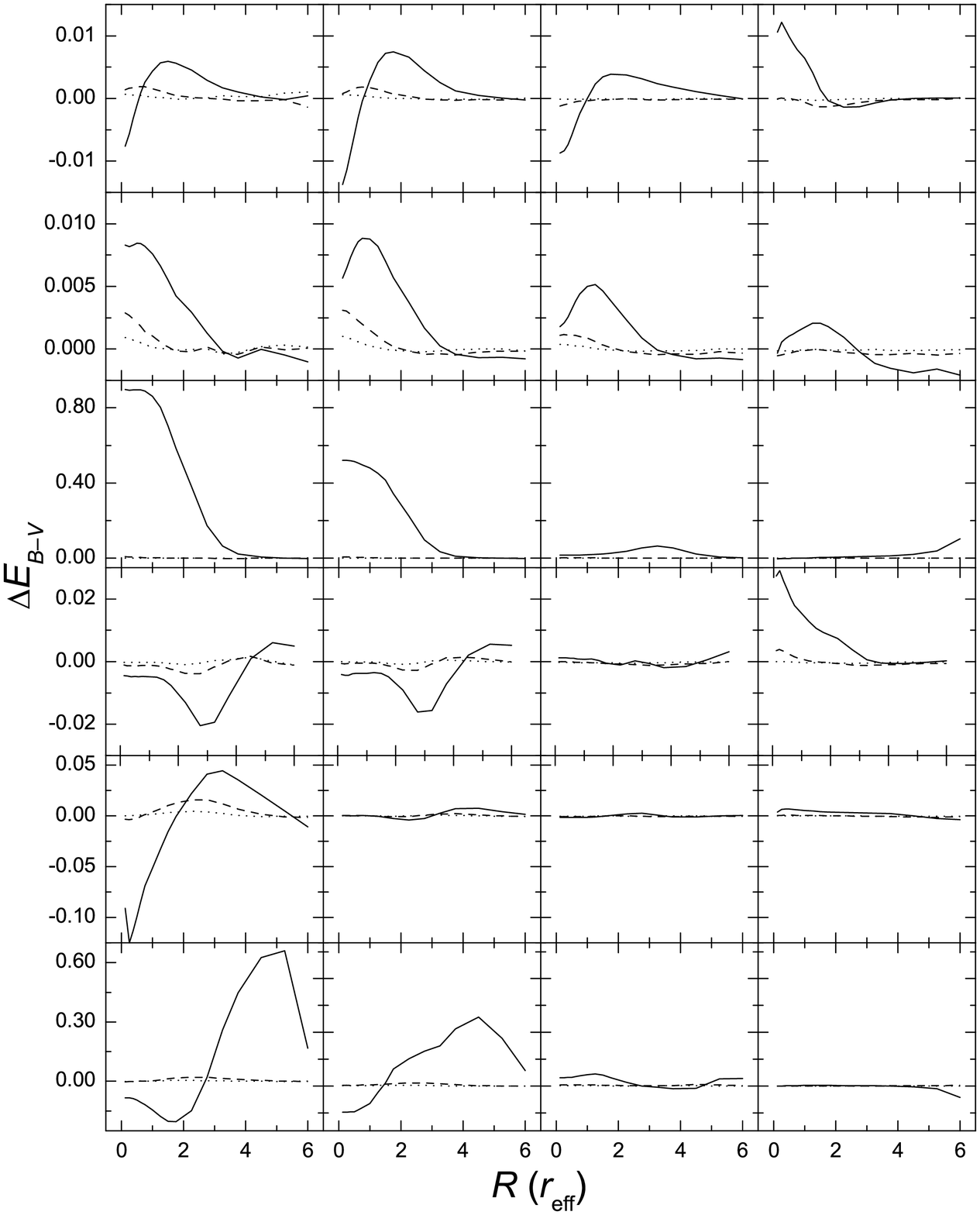,width=124mm,angle=0,clip=}}
\captionb{1}{Differences in
$E_{B-V}$ photometric profile between the models computed with and
without iteration scaling approximation (ISA). The panel rows
correspond to the model groups S1--S6 (starting from top), panel
columns -- to the model galaxy inclinations 0\degr, 50\degr,
80\degr \ and 90\degr \ (left to right). Solid, dashed, dotted
and dash-dotted lines correspond to differences between the M8
model on one side and the  M1, M2, M3 and M5 models on other side.
The aperture radius is given in the effective disk scale-length
$r_{\rm eff}$ units.}
\end{figure}

\begin{figure}
\centerline{\psfig{figure=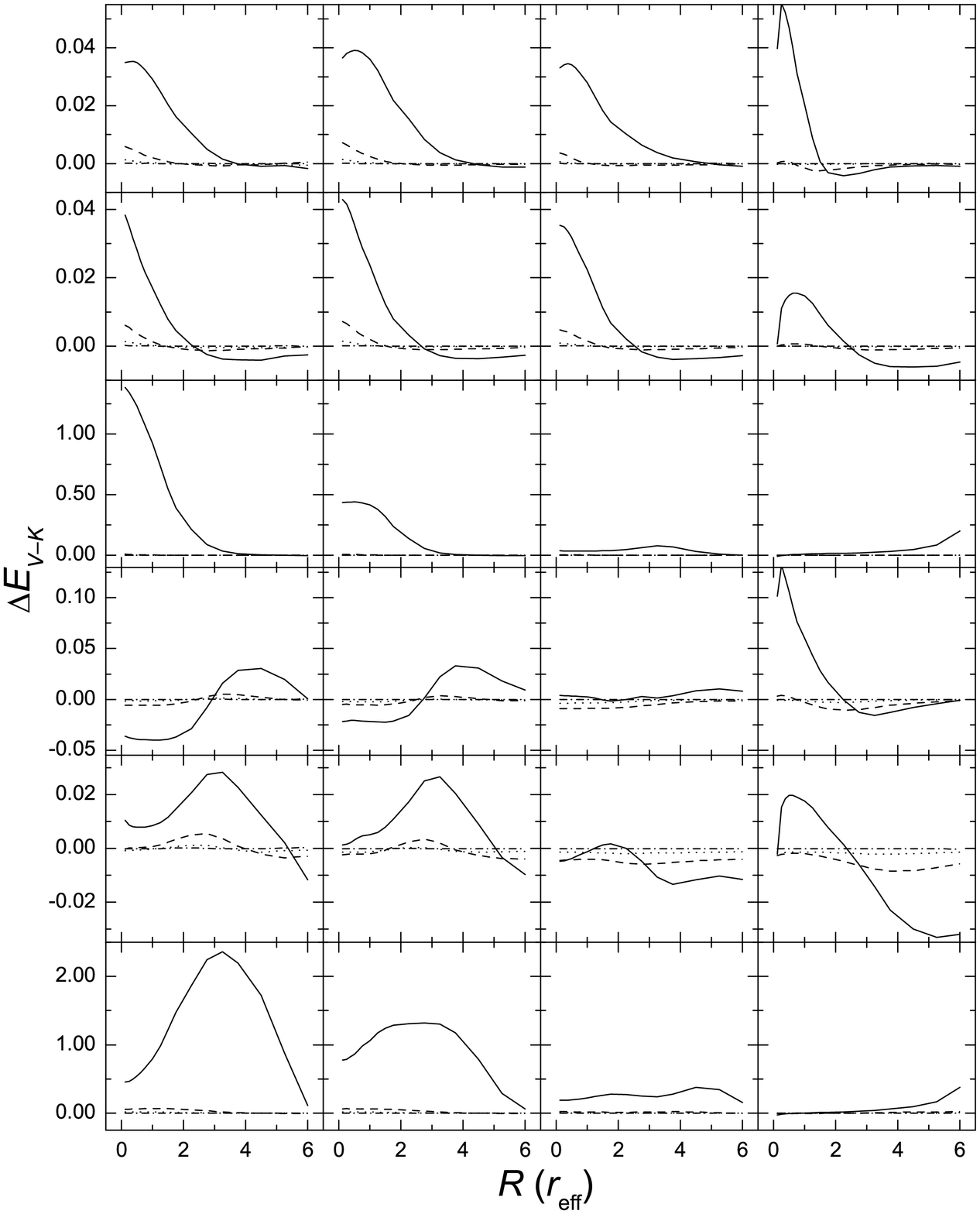,width=124mm,angle=0,clip=}}
\captionc{2}{The same as in Fig. 1, but for the 
$E_{V-K}$ photometric profile.}
\end{figure}

\begin{figure}
\centerline{\psfig{figure=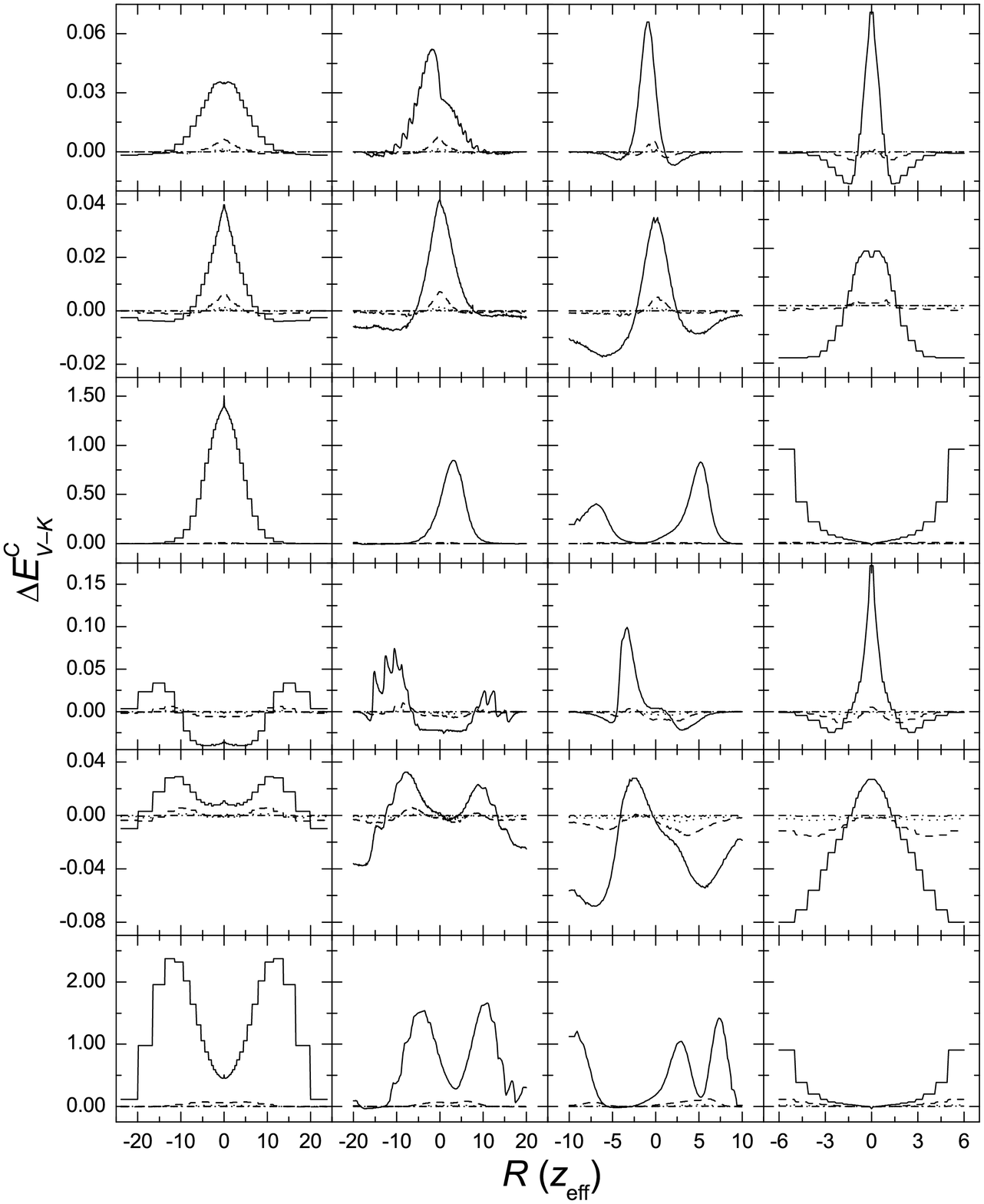,width=124mm,angle=0,clip=}}
\captionb{3}{Differences in the color excess
$E_{V-K}$ minor axis cross-section between the models computed with
and without ISA. Panel rows correspond to the model groups S1--S6
(starting from top), panel columns -- to the model galaxy inclinations
0\degr, 50\degr, 80\degr \ and 90\degr \ (left to right).
Solid, dashed, dotted and dash-dotted lines correspond
to differences between the M8 model from one side and
the M1, M2, M3 and M5 models from other side. The distance
from the image center is given in the effective
disk scale-height $z_{\rm eff}$ units.}
\end{figure}

We compare synthetic multiaperture photometry of the models,
computed using ISA, to the results obtained for exact solution as
$\Delta E_{B-V}^{Mi} = E_{B-V}^{M8} - E_{B-V}^{Mi}$ (Figure 1) and
$\Delta E_{V-K}^{Mi} = E_{V-K}^{M8} - E_{V-K}^{Mi}$ (Figure 2).
Additionally, for all \hbox{models} we have computed difference of color
excess $E_{V-K}$ of the image minor axis cross-sections as
$\Delta E_{V-K}^{C,Mi} = E_{V-K}^{C,M8} - E_{V-K}^{C,Mi}$
(Figure 3).

It can be seen, that approximation-induced errors in all photometric
profiles for the model M1 in groups S1, S2, S4 and S5, while having a
complex radial dependence, remain tolerable for all inclination angles.
In contrast, model groups S3 and S6 display large differences between
models M1 and M8, with the error decrease at high inclinations.  This
behavior at high inclinations is also common for the errors shown by
groups S2 and S5, which, on the average, also show the smallest errors
in low (S1--S3) and high (S4--S6) opacity groups, respectively.

The first-order ISA-induced errors affect most of the minor axis
cross-sections for all considered wavelengths in all models, the
smallest deviation again is observed for groups S2 and S5.  Except for
groups S3 and S6, the deviations of the minor axis cross-section
increase with the increasing inclination angle.

The ISA of higher order produce much better results, M2 being already
sufficiently close to M8, however, in some cases the M3 approximation is
necessary to achieve  acceptable errors.  It can also be noted, that
the error distribution for photometric profiles of $A_V$ and $E_{V-K}$
is identical.

The original ISA method of KB87 seems to successfully reproduce
photometric profiles of the exactly computed models only when the dust
distribution geometrically coincides within the distribution of stellar
population.  The approximation is generally good at high inclinations
even for large optical depths, however, the deviation becomes noticeable
already at 80\degr.  The minor axis cross-section of a model image is
affected more severely and this might lead to erroneous determination of
the geometric parameters of stellar populations for edge-on galaxies
using a comparison with the spectrophotometric models.  The success of
the ISA of higher-orders does not exhibit any significant dependence on
the values of $g_\lambda$ parameters or the model optical depth since
similar results are obtained for all considered passbands.

When determining the required order of ISA it is of prime importance to
fully account for the radiative energy redistribution in the course of
scattering from the primary (stars) to the secondary (dust) sources.
Therefore in most cases, except when the dust and stars have the same
distribution, it is necessary to compute exactly at least the first two
scattering iterations.  Our tests show that the computation of the first
three scattering iterations is generally enough to reach a numerical
accuracy which is undistinguishable from the model computed exactly, and
this decreases the computational time up to five times.

In extreme cases, when the primary and secondary light sources are fully
separated (e.g., dusty shell surrounding dust-free stellar cluster),
more iterations seem to be necessary.  In these cases the applicability
of ISA should be properly investigated.  The usefulness of ISA would
increase considerably if we were able to establish criteria, determining
whether a given number of iterations is sufficient.

\vskip7mm

ACKNOWLEDGMENT.\ This work has been supported by the Lithuanian
State Science and Studies Foundation.

\newpage

\References

\refb Baes M., Dejonghe H. 2001, MNRAS, 326, 722

\refb Henyey L., Greenstein J. 1941, ApJ, 93, 70

\refb Kylafis N., Bahcall J. 1987, ApJ, 317, 637

\refb Laor A., Draine B. 1993, ApJ, 402, 441

\refb Li A., Greenberg M. 2003, in {\it Solid State Astrochemistry},
NATO Science Series II, eds.  V. Pirronello, J. Krelowski \& G.
Manic{\`o}, Kluwer Academic Publishers, Dordrecht, p. 37

\refb Pascucci I., Wolf S., Steinacker J., Dullemond C., Henning Th.,
Niccolini G., Woitke P., Lopez B. 2004, A{\&}A, 417, 793

\refb Semionov D., Vansevi{\v c}ius V. 2002, Baltic Astronomy, 11, 537

\refb Semionov D. 2003, {\it Spectrophotometric Evolution of Dusty Disk
Galaxies}, Inst. of Theoretical Physics and Astronomy, Vilnius, PhD
thesis

\refb Semionov D., Vansevi{\v c}ius V. 2005a, astro-ph/0501146

\refb Semionov D., Vansevi{\v c}ius V. 2005b, Baltic Astronomy, 14, 235
(this issue)

\end{document}